\documentclass[conference]{IEEEtran}

\usepackage{cite}
\usepackage{amsmath,amssymb,amsfonts}
\usepackage{algorithmic}
\usepackage{graphicx}
\usepackage{textcomp}
\def\BibTeX{{\rm B\kern-.05em{\sc i\kern-.025em b}\kern-.08em
    T\kern-.1667em\lower.7ex\hbox{E}\kern-.125emX}}
\begin{document}

\title{Learning more from crossing levels: Investigating agility at three levels of the organization %\\(CSCI-ISSE: philosophical short paper)
}

\author{\IEEEauthorblockN{Lucas Gren}
\IEEEauthorblockA{Chalmers University of Technology and The University of Gothenburg\\
Gothenburg, Sweden 412--92\\
lucas.gren@cse.gu.se}

}

\maketitle
\begin{abstract}
Scholars have tried to explain how organizations can build agile teams by only looking at one level of analysis. We argue in this short paper that lessons can be learned from organizational science results explaining variance on three different abstraction levels of organizations. We suggest agility needs to be explained from organizational (macro), the team (meso), and individual (micro) levels to provide useful and actionable guidelines to practitioners. We are currently designing such studies and hope that they will eventually result in validated measurements that can be used to prevent companies from investing in the wrong areas when trying to move towards more agility.

\end{abstract}

\begin{IEEEkeywords}
agile, software engineering, software development, project management
\end{IEEEkeywords}

\section{Introduction}
With the paradigm shift in software development toward more flexible and agile development processes, comes a need for more adaptive leadership \cite{cohn2004situational}. However, this adaptive leadership always happens in a context, and this context is influenced by factors at different abstraction levels of the organization \cite{hackman2003learning}. Some studies have shown that agility needs to be present at a strategic level as well to fully be implemented and for the organization to gain the intended advantages \cite{scrumorbeing}. Also, in industry, many agile transitions fail because of lack of team capabilities \cite{chow2008survey}. Leadership can be viewed as a person's ability to influence other individuals \cite{2009tei}, but how people are influenced is also dependent on the maturity of teams, what phase the project is in, and the organization as a whole. 

From a team perspective, measurement is essential to evaluate progress \cite{kozlowski2006enhancing}. However, one of the most significant challenges is to measure at the right abstraction level \cite{hackman2003learning}. Hackman \cite{hackman2003learning} divides levels of abstraction into three levels, namely micro, meso, and macro. When researching the creation of agile teams, we want to focus on group-level development, hence, the meso level. However, explanations to variance we observe might very well come from the macro (organizational) level or the micro (individual) level. Therefore, if we want to understand agility and how such cultural change affects a software development organization, we should, preferably, look at all three levels. Therefore, we want to conduct research on all three levels simultaneously and urge other scholars to follow suit. 

\section{Lessons learned from crossing levels in other fields}
Hackman \cite{hackman2003learning} presents some examples of when another level of analysis could reveal where the variance was hiding in some of his organizational studies. When investigating how airline cockpit crews deal with problems and issues before they became severe, they found no explanations across teams when measuring team performance effectiveness, which comprised the design of flying tasks and the composition of the crews. Looking one level down, they also investigated personal leadership styles, which also failed to explain any differences across airlines. The differences across airlines were purely organizational. Together, cockpit technology, the regulatory environment, and the culture of flying, completely constrained any crew from making any difference in relation to their performance. Not even accidents change the way these airlines assemble crews, but instead always result in technological fixes. In relation to agile teams, trying to explain why some teams manage to self-organize by only looking at team-level differences, might as well be futile. The properties of each organization might be the critical discontinuing factor that needs fixing before any team can move towards more agility. Also, training staff in agile leadership would then also not give any tangible results. Some scholars in the agile software engineering research community have suggested which strategic factors need to be in place for an agile transition (see, e.g.\ \cite{sidky}), however, to get the full picture we need to investigate the micro, meso, and macro levels at the same time. 

Next, we will present the three levels we investigate at the moment and, after that, we discuss potential outcomes of such research. 

\section{Situational leadership}\label{sub:situational}
\begin{figure}
\centerline{\includegraphics[width=80mm]{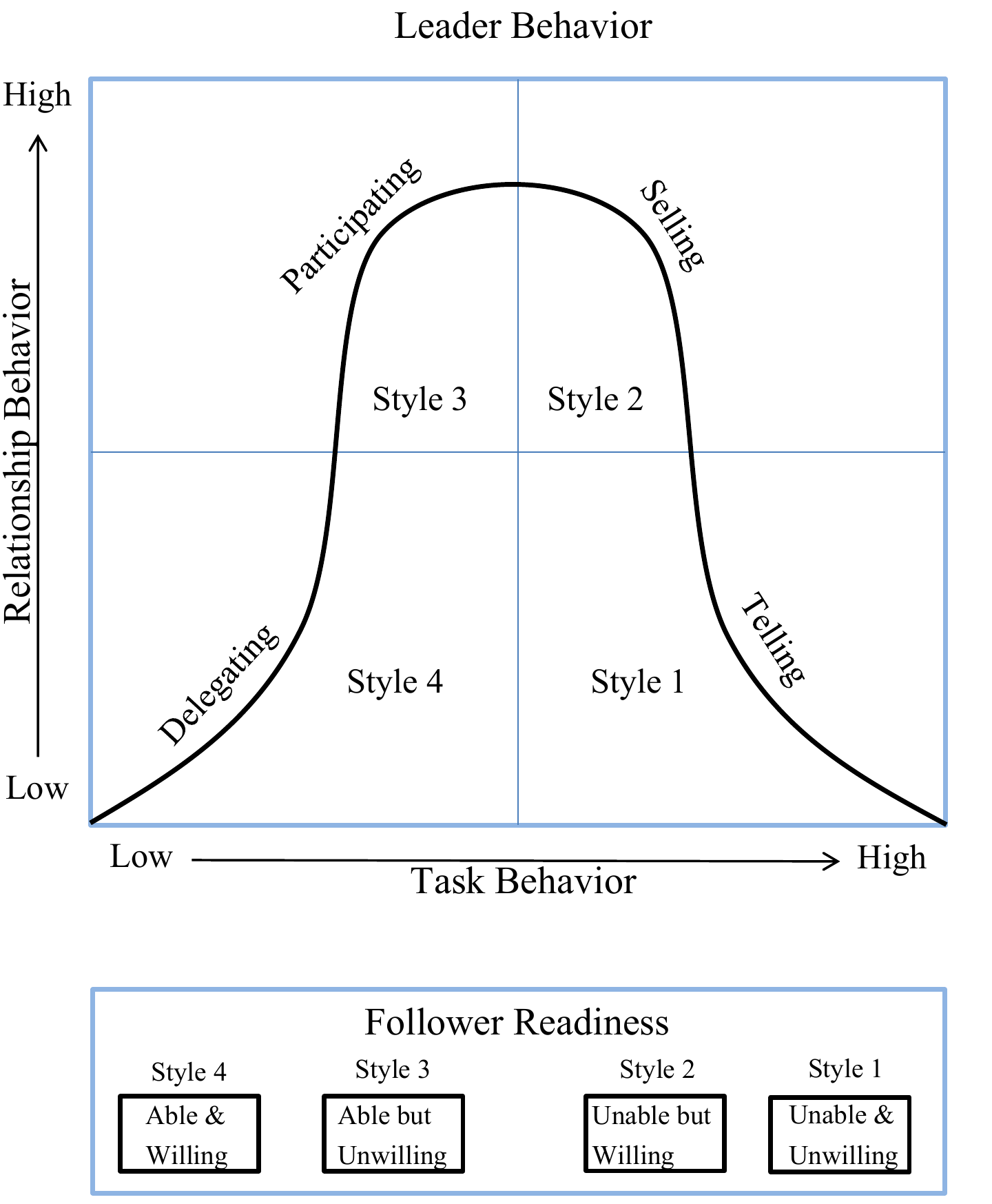}}
\caption{Situational leadership (adopted from \cite{hersey}).}
\label{fig:situationalleadership}
\end{figure}

Instead of finding an optimal leadership style, Hersey et al. \cite{hersey} suggested already in the seventies that a leader much adapt and change the style depending on the group. This model consists of maturity levels of group members, but also a balance between relation- and task-oriented behaviors. A leader should act differently depending on the needs of the group. Modern organizational psychology scholars also advocate a dynamic team leadership adapted to emerging needs even in the same situation \cite{2009tei}. The steps suggested by Hersey et al. \cite{hersey} are illustrated in Figure~\ref{fig:situationalleadership}. 

\section{An integrated model of group development}\label{sub:groupdev}
\begin{figure}
\centerline{\includegraphics[width=90mm]{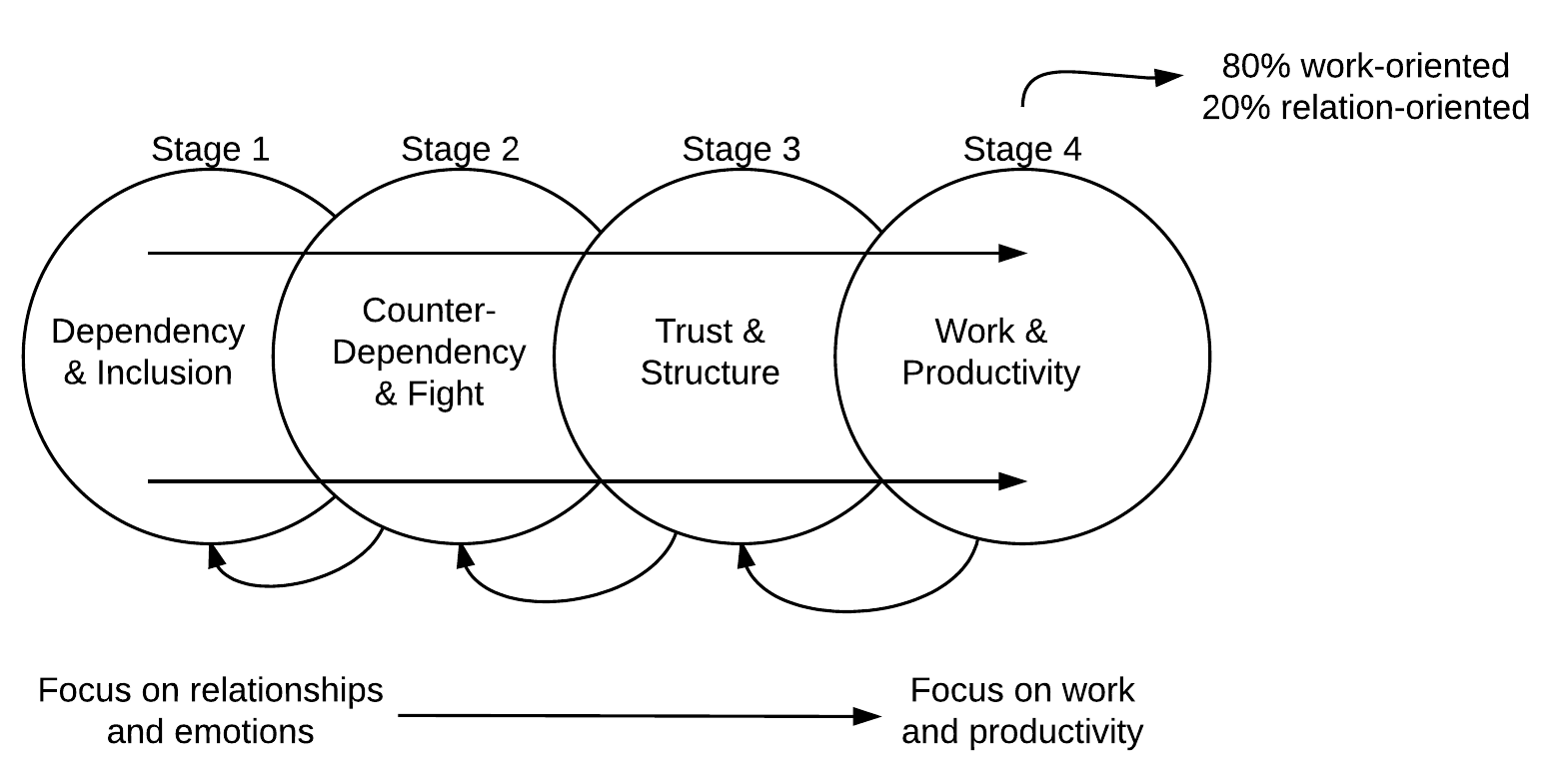}}
\caption{The Group Development Stages. Adopted from~\cite{wheelan2012}.}
\label{fig:groupstages}
\end{figure}

Due to many years of research on groups, there are a diversity of group development models. Even though the models are somewhat different, there is e a reoccurring pattern. A popular integrated model was suggested by Tuckman~\cite{tuckman} in 1965 with the phases; Forming, Storming, Norming, and Performing. A bit more recently, Wheelan \cite{wheelan} created a model called the Integrated Model of Group Development (or IMGD) that can more or less be translated into the stages created by Tuckman~\cite{tuckman}. The stages are shown in Figure~\ref{fig:groupstages}. The stages can be compared to those of a human. We first figure out what world we are in (being a child), then we question the rules and their existence (adolescence), and we eventually somewhat find our place in this world and can focus more on how to develop and mature in life. It has been shown that, on an overall level, human groups go through similar stages \cite{wheelan}.

%The Integrated Model of Group Development presents four different temporal stages that all groups go through on their journey to becoming a mature high performing team. These stages are shown in Figure~\ref{fig:groupstages} and are straight-forward because we have an intuitive understanding and know that we act differently with people we know and people we do not know. The developmental levels of groups can be compared to that of a human. We first figure out what world we are in (being a child), then we question the rules and their existence (adolescence), and we eventually somewhat find our place in this world and can focus more on how to develop and mature in life \cite{2003hop}. It has been shown that, on an overall level, human groups go through similar stages \cite{wheelan}.

\paragraph{Stage 1: Dependency and inclusion}
The first stage is categorized by three main areas; concerns about safety and inclusion, member dependency on the designated leader, and a wish for order and structure. The group is supposed to become organized, capable of efficient work, and achieve goals, so the first state must have a purpose in getting there \cite{wheelan}. 

\paragraph{Stage 2: Counter-dependency and fight}
When the group safely navigated through the previous stage, the group members have gained a sense of loyalty. When people feel safer, they will dare to speak up and express opinions that might not be shared by all members. The second stage of a group's development is, therefore, a conflict phase where a fight is a must to create clear roles to be able to work together constructively. The members have to go through this to be able to trust each other and the leader \cite{wheelan}.

\paragraph{Stage 3: Trust and structure}
The third stage is a structure-developing phase where the roles are based on competence instead of striving for power or safety. Communication will be more open and task-oriented. The third stage of group development is characterized by more mature negotiations about roles, organization, and processes \cite{wheelan}. 

\paragraph{Stage 4: Work and productivity}
The fourth and final stage (excluding the termination phase) is when the group wants to get the task done well at the same time as the group cohesion is maintained over an extended period. The group also focuses on decision-making and encourages task-related conflicts. This stage is a time of intense productivity and effectiveness, and it is at this stage the group becomes a team \cite{wheelan}.

\section{Project and group life-cycles}\label{sec:projectlifecycle}
It is more and more common to work in the form of projects within organizations, and a project goes through a set of stages that can be described as Idea, Planning, Execution, and Termination \cite{ricciardi}. The first stage (the idea stage) is when the idea comes to place, and the company realizes that a project is needed around a specific goal.  The planning stage comprises detailed planning, budgeting, scheduling, recruitment, and procurement. The execution stage is when the main project work gets done, and in the termination stage, the work decreases and the results are delivered to the customer. 

\begin{figure}
\centerline{\includegraphics[width=90mm]{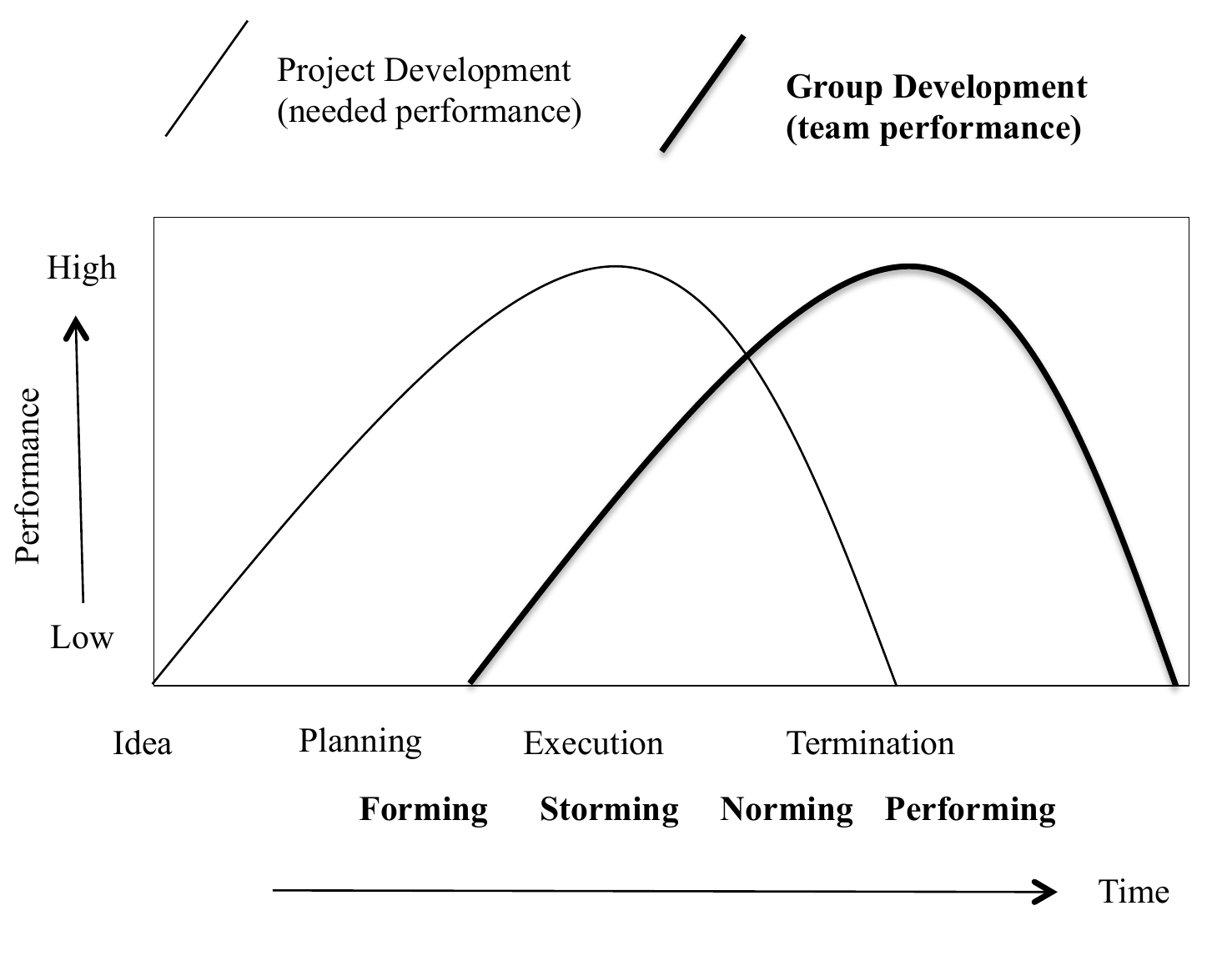}}
\caption{Project and group development stages (adopted from \cite{ricciardi}).}
\label{fig:projectstages}
\end{figure}

The group development life-cycle described in Section~\ref{sub:groupdev} and the project's life cycle have a mutual effect on each other, and the problem is that the group's development and the project's development are rarely synchronized. Therefore, the group members could avoid sharing their opinions in the project planning stage, because the group is, psychologically, in the Forming stage. Also, the group might as well be in the conflict (Storming) stage during project execution, which is when the team's performance needs to be at its peak. As the cycle moves along, the team might be starting to perform at its best when the project terminates \cite{ricciardi}. All these effects, are, of course, suboptimal (see Figure~\ref{fig:projectstages}).

\section{Organizational maturity}\label{sec:sililarities}
\begin{figure*}
\centerline{\includegraphics[width=140mm]{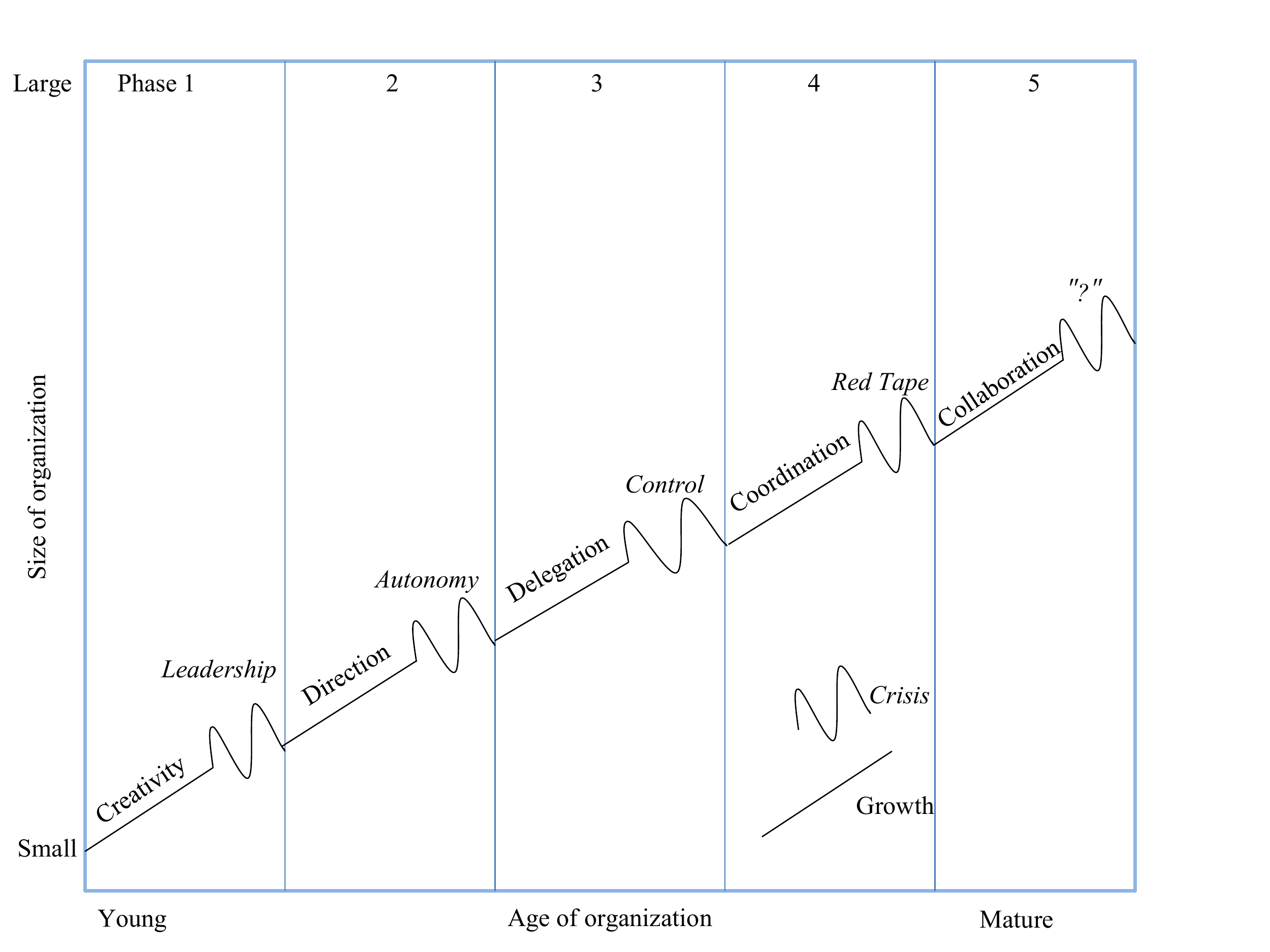}}
\caption{Organizational development stages (adapted from \cite{greiner1972}).}
\label{fig:orgstages}
\end{figure*}

In the group development model in Section~\ref{sub:groupdev}, the project development model, presented in the previous section, and the situational leadership model in Section~\ref{sub:situational}, the phases are divided into a formation stage, a crisis stage, a norming stage, and a work stage. These can also be compared to, for example, Greiner's \cite{greiner1972} model for growing organizations (see Figure~\ref{fig:orgstages}). A new organization that is starts with the entrepreneurial phase that is characterized by growth through creativity. The manager is here individualistic, creative and an entrepreneur. At the end of Phase 1, the organization has a leadership crisis. The following phase is the collective phase where growth is managed through directives that often come from the leader. At the end of this phase, the organization goes through a crisis of autonomy. Phase 3 is a phase of formalization where growth is managed through delegation. Total delegation and autonomy is given from the leader but ends with a crisis of control. The development phase (Phase 4) is to grow by coordination. The leader acts as a watchdog and this phase often ends up in a crisis of bureaucracy (sometimes called `red tape'). The final stage is recognized through team-oriented work and interpersonal skills where learning and innovation are present. While we realize that this publication is less scientific, it largely overlaps with other findings in organizational dynamics (see, e.g.\ \cite{adizes1979organizational}). These phases are very similar to the group development stages, human development, situational leadership, and project development. The challenge is to be aware of these and synchronize them carefully. 

\section{Discussion}\label{sec:disc}

The fact that groups have different needs over time could be connected to leadership research that focuses on that different leadership is needed in different contexts depending on what the group and group-members need. The situational leadership model \cite{hersey} includes maturity levels of the group members, but also that balance is needed between relation- and task-oriented leadership behavior. They present the four different styles `telling', `selling,' `participating,' and `delegating' that are somewhat translatable to the group development stages. With more immature groups, telling and selling are needed approaches for the leadership to be successful, while at more mature stages, participating and, finally, delegating are more effective styles, since the group can self-organize. This means that the agile practices need to be implemented using a different leadership style depending on the maturity level of the group. In the agile method Scrum, we find descriptions of `agile leadership' as being facilitating instead of directing \cite{schwaber,chin}. This works well in a mature group, but if that is not the case, the leader will need to behave differently to move the group forward. That is why situational leadership adapted to the group development stages needs to be incorporated into software engineering processes, and if they are not, leaders and managers will wrongfully try to follow a method that could be hindering the progress of that specific team.

Connecting the three levels of agility to the three more general abstraction levels of organizational theory is needed to fully understand agility. As mention, there is some evidence showing that agility in its broader sense is required at all levels of an organization to reach the intended increase in productivity and it is possible to change the practices on a more superficial level without the cultural change \cite{scrumorbeing}. We, therefore, need the whole organization to be on board with our agile transition, which is not very surprising, but difficult in practice. Returning to Greiner's \cite{greiner1972} model of growing organizations (see Figure~\ref{fig:orgstages}), we can see that small organizations, like start-ups, are agile by definition, i.e.\ they do not have any extensive overhead processes to satisfy when making decisions or negotiating with customers, and are characterized by creativity. However, they will have a leadership crisis sooner or later when the organization gets too large, with the reason being that the founder is then unable to obtain an overview and have control over all operations in the organization. This reasoning provides an explanation to why larger companies can not, and should not, function like small start-ups. However, there are different approaches to growing an organization than the classical command-and-control paradigm, i.e.\ to instead: `trust the collective intelligence of the system' \cite{laloux2014roa}; however, such organizations are still rare. We also believe it is too early to know the potential of such approaches at a much larger scale since they are more of an exception than a rule today, but we see potential in extending the agile methods with ideas of entirely autonomous teams.

In the software engineering community, a large focus has been on internal process maturity. However, it is also cumbersome to only look at the process maturity, like CMMI (Capability Maturity Model Integration) or the ISO\slash IEC 15504 SPICE (Software Process Improvement and Capability Determination), since we want agile value-driven organizations that use agile practices to implement agile principles. In addition, process maturity models are based on building customer trust by process infrastructure instead of working software and customer participation \cite{turner20022}. The strategists (managers\slash leaders) and the employees of an organization need to set the vision according to the organization's purpose of existence in alignment with the agile principles (the cultural change) and then select agile practices to support that journey \cite{williams} in relation to all three abstraction levels of the organization, i.e.\ the micro, meso, and macro perspectives, namely the organizational, team\slash project, and individual levels. Understanding more about these interactions could increase the predictability of when agile transformation efforts succeed or fail and provide explanations for why.

%\section*{Acknowledgment}

%\section{Conclusion and Future Work}

\bibliographystyle{IEEEtran}

\bibliography{references}

\end{document}